\documentclass{iopart}
\usepackage{graphicx}
\begin{document}

\title{Generalized integral fluctuation relation with feedback control for diffusion processes}
\author{Fei Liu and Hongcheng Xie}
\address{School of Physics and Nuclear Energy Engineering, Beihang University, Beijing 100191, China}
\ead{feiliu@buaa.edu.cn}
\author{Zhiyue Lu}	
\address{Chemical Physics Program, University of Maryland, College Park, College Park, MD 20742}

\begin{abstract}
We extend a generalized integral fluctuation relation in diffusion
processes that we obtained previously to the situation with
feedback control. The general relation not only covers existing
results but also predicts other unnoticed fluctuation relations.
In addition, we find that its explanation of time-reversal
automatically emerges in the derivation. This interesting
observation leads into an alternative inequality about the
entropy-like quantity with an improved lower bound. Two
feedback-controlled Brownian models are used to verify the result.
\end{abstract}
\submitto{\JPA}

\section{Introduction}
In the past two decades, an important progress of nonequilibrium
physics  is the discovery of a variety of fluctuation
relations~\cite{Evans,EvansSearles,Gallavotti,Kurchan,Lebowitz,Bochkov77,JarzynskiPRL97,
JarzynskiPRE97,Crooks99,Crooks00,HatanoSasa,Maes,SeifertPRL05,Speck,Kawai,EspositoPRL10}.
These exact relations about the statistics of entropy production
or dissipated work have greatly deepened our understanding of the
second law of thermodynamics and nonequilibrium physics of small
systems. Recently, these fluctuation relations were also extended
into other fields, {\it e.g.}, the thermodynamics of
information-processing systems~\cite{Leff,CaoEntropy12}. For
instance, Sagawa and Ueda~\cite{SagawaPRL10} obtained a
generalized Jarzynski equality with feedback control,
\begin{eqnarray}
\label{orgSagawaUedaEquality} \langle e^{-\beta
W_{diss}-I}\rangle=1,
\end{eqnarray}
where $\beta$ is the inverse temperature of heat bath, $ W_{diss}$
is the dissipated work~\cite{JarzynskiPRL97,JarzynskiPRE97}, and
$I$ is the mutual information. This intriguing work equality leads
into an inequality,
\begin{eqnarray}
\label{SagawaIneq}\beta\langle W_{diss}\rangle \ge -\langle
I\rangle,
\end{eqnarray}
which agrees with the second law of information
thermodynamics~\cite{CaoPRE09}. The validity of these relations
has been verified by the single colloidal particle
experiment~\cite{Toyabe}. Very recently, Abreu and
Seifert~\cite{AbreuPRL12} extended Sagawa-Ueda's
equality~(\ref{orgSagawaUedaEquality}) into genuine nonequilibrium
processes in the master equation systems. Analogous equalities and
inequalities with feedback control were obtained for the excess
entropy production and total entropy production. Other more
general relations were also
reported~\cite{KunduPRE12,LahiriJPA12,SagawaPRL12}.

As we mentioned at the beginning, in classical systems there
are various fluctuation relations. The results of Sagawa and Ueda
~\cite{SagawaPRL10}, and Abreu and Seifert~\cite{AbreuPRL12} have
implied the plausibility of introducing feedback control into
those relations. Moreover, we note that in the past few years
there were quiet a few efforts of unifying the fluctuation
relations~\cite{Chernyak,Taniguchi,Chetrite,LiuFPRE09,LiuFJPA09,LiuFJPA10}.
Hence, it shall be more desirable if these various relations with
feedback control could be as well unified under a single formula.
Here we show that the unification does exist for the continuous
diffusion processes. Instead of starting from the conventional
detailed fluctuation theorem~\cite{SagawaPRL10,
KunduPRE12,LahiriJPA12,HorowitzPRE10,SagawaPRE12}, our theory is
based upon the generalized fluctuation integral relation
(GIFR)~\cite{LiuFPRE09,LiuFJPA10} that we proposed earlier. Its
advantage is that the GIFR does not need to explicitly define {\it
time-reversal}. Hence, it is capable to unify the various
fluctuation relations relying on distinct time-reversals, {\it
e.g.}, that of the Hatano-Sasa
equality~\cite{HatanoSasa,EspositoPRL10}.

The organization of the work is as follows. In
Sec.~\ref{reviewofGIFR} we briefly review the GIFR in the
diffusion processes. In Sec.~\ref{GIFRfeedbacksection} we extend
it by including feedback control. In Sec.~\ref{discussion} we
first show that the generalized relation not only covers the
existing results but also gives other previously unnoticed
equalities. Then we point out its time reversal explanation and
obtain an improved inequality about the entropy-like quantity. In
the end, we use two feedback-controlled Brownian particles to
verify this result. Section.~\ref{Conclusion} is a summary of this
work. Appendix A extends the GIFR with feedback control in the
diffusion processes into the continuous time master equations with
discrete states. Appendix B explains the Bochkov-Kuzovlev equality
(BKE) in overdamped Brownian motion from the point of view of the
GIFR.

\section{Overview of GIFR in diffusion processes}\label{reviewofGIFR}
Consider a general $N$-dimension ($N$-d) stochastic system with
variables $x$$=$$\{x_i\}$, $i$$=$$1$,$\cdots$,$N$.  The dynamics
of the system is described by the stochastic differential
equations (SDE)~\cite{Gardiner,Risken}
\begin{eqnarray}
d{x}(t)={A}({x},t)dt + \sqrt{{B}({x},t)}d{W}(t), \label{SDE}
\end{eqnarray}
during time interval $(t_0$, $t_f)$, where $d{W}$ is an $N$-d
Wiener process, ${A}$$=$$\{A_i\}$ denotes a $N$-d drift vector,
and $ \sqrt{{B} }$ is the square root of a $N$$\times$$N$ positive
definite and symmetric diffusion matrix ${B}$. In the work we
follow the Ito's convention for the SDE. Rather than directly
solving~(\ref{SDE}), one usually converts it into the evolving
equation of the probability density function (pdf) $\rho({x},t)$
of the system, {\it i.e.}, the Fokker-Planck equation, $\partial_t
\rho$$=$${\cal L}({x},t)\rho$. The Fokker-Planck operator is
\begin{eqnarray}
\label{FPoperator} {\cal L}({x},t)=-\partial_{x_i} A_i({x},t)
+\frac{1}{2}\partial_{x_i}\partial_{x_l} B_{il}({x},t),
\end{eqnarray}
where the Einstein's summation convention is used. We proved that
the ${x}$-integral,
\begin{eqnarray} \label{Rfunction}\fl\hspace{1cm} \int dx
\varrho(x,t') R(t_f|{x},t')=\int dx \varrho(x,t')
\hspace{0.1cm}^x\left\langle e^{-\int_{t'}^{t_f}{\cal
J}[\varrho,{S}]({x}(\tau),\tau)d\tau }O(x (t_f))\right\rangle,
\end{eqnarray}
is a $t'$-invariant
($t_0$$\le$$t'$$\le$$t_f$)~\cite{LiuFPRE09,LiuFJPA10}, where
$\varrho({x},t')$ is an arbitrary pdf, $O({x})$ is an arbitrary
function, and the average $^{x }$$\langle$$\cdots$$\rangle$ is
over all stochastic trajectories starting from $x$ at time $t'$
and following~(\ref{SDE}). In the above equation we defined
\begin{eqnarray}
\label{Jfunctional} \fl\hspace{0.5cm} {\cal J}[\varrho,{S}]
=\varrho^{-1}\left[\left({\cal
L}-\partial_\tau\right)\varrho+2\partial_{x_i}S_i +2\varrho^{-1}
S_iB^{-1}_{il}
S_l\right]+2\varrho^{-1}S_iB^{-1}_{il}\left(\dot{x}_l-A_l \right),
\end{eqnarray}
where the dot denotes the time derivative $d/d\tau$,
$S$$=$$\{S_i\}$ is a $N$-d vector field which is either zero at
the boundary of the system or is periodic if the system is
periodic. Particularly, if we let $t'$=$t_0$,~(\ref{Rfunction})
implies an equality
\begin{eqnarray}
\left\langle e^{-\int_{t_0}^{t_f}{\cal
J}[\varrho,{S}]({x}(\tau),\tau)d\tau
}O(x(t_f))\right\rangle_{\varrho_0}=\langle
O\rangle_{\varrho_{f}},\label{GIFR}
\end{eqnarray}
where the subscripts $\varrho_0$ and $\varrho_f$ indicate that the
averages are done over $\varrho({x},t_0)$ and $\varrho({x},t_f)$,
respectively. Since (\ref{GIFR}) is a general mathematic
identity and can cover several important fluctuation
relations~\cite{Bochkov77,JarzynskiPRL97,HatanoSasa,SeifertPRL05}
by choosing specific $\varrho$ and $S$, we named it the GIFR.

The function $R(t_f|{x},t')$ has an interesting time-reversal
explanation. Assume the variables ${x}$ of the stochastic system
to be even or odd according to their rules under time reversal:
$x_i$$\to$$x_i$ is even and $x_i$$\to$$-x_i$ is odd; in
abbreviation $x_i$$\to$$\tilde{x}_i$$ =$$\varepsilon_i x_i$ with
$\varepsilon_i$$=\pm1$. We found~\cite{LiuFPRE09,LiuFJPA10}
\begin{eqnarray}
\label{Texplanation} R(t_f|{x },t')\varrho({x},t')=\langle O\rangle_{\varrho_f}q({\tilde{x}},s), 
\end{eqnarray}
where $s$$+$$t'$$=$$t_f$, $q({x},s)$ is the pdf of the
time-reversed Fokker-Planck equation, whose drift force and
diffusion matrix are
\begin{eqnarray}
\label{reversedforce}\tilde{A}_i({x }, s)=-\varepsilon_i
A_i({\tilde{x}}, t')+ \frac{2\varepsilon_i}{\varrho({\tilde{x}},
t')}\left[S_i
+\frac{\varepsilon_l}{2}\partial_{x_l}(B_{il}\varrho)\right]({\tilde{x}},
t'),
\end{eqnarray}
and
\begin{eqnarray}
\label{reverseddiffusion} \tilde{B}_{il}({x
},s)=\varepsilon_i\varepsilon_l B_{il}({\tilde{x}},t'),
\end{eqnarray}
respectively, and its initial condition is $O({x
})\varrho({x},t_f)/\langle O\rangle_{\varrho_f}$~\footnote{The
$O(x)$ function is now nonnegative in order to have a probability
interpretation.}. Note that there is no summation over the indice
$i$ here.

The time-reversal explanation leads into two important
consequences. First, the generalized detailed balance relation was
established~\cite{Chetrite,LiuFJPA10} :
\begin{eqnarray}
\label{GDetailbalance}R({x }_2,t_2|{x }_1,t_1)\varrho({x}_1,t_1)=
G({\tilde {x }}_1,s_1|{\tilde {x}}_2,s_2)\varrho({x }_2,t_2).
\end{eqnarray}
where $R({x}_2,t_2|{x }_1,t_1)$ is a specific $R(t_2|x_1,t_1)$
in~(\ref{Rfunction}) with $O({x})$=$\delta({x}$$-$${x}_2)$, and
$G({x}_1,s_1|{x }_2,s_2)$ is the transition probability of the
time-reversed process from earlier time $s_2$$=$$t_f$$-$$t_2$ to
the later` time $s_1$$=$$t_f$$-$$t_1$. Second, applying the
Jensen's inequality to~(\ref{Texplanation}) with $O(x)$$=$$1$, we
obtain an inequality
\begin{eqnarray}
\label{inequality} \left\langle\int_{t_0}^{t_f}{\cal
J}[\varrho,{S}]({x }(\tau),\tau)d\tau \right\rangle_{\varrho_0}\ge
D[\varrho({x},t_0)||
q({\tilde{x}},t_f)],
\end{eqnarray}
where the right-hand side is the relative entropy between the two
pdfs, which is always nonnegative~\cite{Cover}. We name the left
hand side of~(\ref{inequality}) the entropy-like quantity. We did
not present this result in our previous work. Several specific
cases can be found in~\cite{JarzynskiRev11}.
 
\section{GIFR with feedback control}
\label{GIFRfeedbacksection} Rather than considering very general
diffusion processes, in the remaining sections we focus on those
that the time-dependence of their drift vectors and diffusion
matrixes only comes from a set of external control parameters
$\lambda_t$. During the whole process, we assume that there are
$M$ measurements and feedback loops applied on~(\ref{SDE}) at
discrete times $t_k$, $k$$=$$1$,$\cdots$,
$M$~\cite{HorowitzPRE10,SagawaPRE12}. From time $t_0$ to $t_1$,
the system evolves under the predetermined protocol $\lambda^0_t$.
In the following time, at each $t_k$, a physical observable is
measured and its outcome $y_k$ is obtained with pdf
$P(y_k|{x}_{t_k})$. During time interval $(t_k,t_{k+1})$ the
parameters $\lambda_t^k$ vary according to a {\it continuously}
protocol that is uniquely determined by previous outcomes
$\mu_k$=$\{y_1$,$\cdots$,$y_k\}$ up to time $t_k$. We explicitly
indicate this point by notation $\lambda_t^k(\mu_k)$.
Additionally, we let $t_{M+1}$=$t_f$. After completion of one
process, we have a protocol $\Lambda$=$\{$$\lambda_t^0$,$\cdots$
$\lambda_t^M(\mu_M)$$\}$. With these notations, we
extend~(\ref{Rfunction}) into the case with feedback control:
\begin{eqnarray}
\label{RfunctionFeedback} \fl^{{x}_0}\left\langle
e^{-\sum_{k=0}^M\int_{t_k}^{t_{k+1}}{\cal J}[\varrho,{S}]({x
}(\tau),\tau,\lambda^k_\tau)d\tau -J-I}O({x}(t_f))\right\rangle
\varrho({x }_0,t_0)=\langle\langle O\rangle_{\varrho_f }
q_{\tilde{\Lambda}}({\tilde{x}}_0,t_f)\rangle_{\mu_M},
\end{eqnarray}
where the averages are done over all possible outcomes and
trajectories, $\varrho$ and $S$ may or may not be functions of the
control parameters, and the trajectory-dependent terms $J$ and
$I$~\cite{HorowitzPRE10} are
\begin{eqnarray}
\label{jumpprotocols} J=\ln\prod_{k=1}^{M}\frac{
\varrho(x(t_k),t_k,\lambda^{k-1}_{t_k})}{\varrho(x(t_k),t_k,\lambda^{k}_{t_k})},
\end{eqnarray}
and
\begin{eqnarray}
\label{tjinformation} I=\ln\prod_{k=1}^{M}\frac{
P(y_k|x(t_k))}{P(y_k|\mu_{k-1})},
\end{eqnarray}
respectively. We have denoted $P(y_1)$$=$$P(y_1|\mu_0)$. On the
right-hand side of~(\ref{RfunctionFeedback}) $q_{\tilde{\Lambda}
}({\tilde{x}}_0,t_f)$ is the pdf of the time-reserved process~(7)
at time $t_f$ under the specific time-reversed protocol
$\tilde{\Lambda}$=$\{$$\tilde{\lambda}^M_s(\mu_M)$,$\cdots$,$\tilde{\lambda}^0_s$$\}$
and $\tilde{\lambda}^k_s(\mu_k)$$=$$\lambda^k_{t_f-s}(\mu_k)$. The
$J$-term~(\ref{jumpprotocols}) arises from possible discontinuity
of the control parameters at the measurement times. Such type of
protocols was often used in
modelling~\cite{CaoEntropy12,AbreuPRL12,SagawaPRE12,FeitoPRE09,LopezPRL08,AbreuEPL11}.
Obviously, if we integrate the two sides
of~(\ref{RfunctionFeedback}) on $x_0$, we have the GIFR with
feedback control:
\begin{eqnarray}
\label{GIFRFeedback} \left\langle e^{-
\sum_{k=0}^M\int_{t_k}^{t_{k+1}}{\cal J}[\varrho,S]({x
}(\tau),\tau,\lambda^k_\tau)d\tau
-J-I}O(x(t_f))\right\rangle_{\varrho_0}=\langle\langle
O\rangle_{\varrho_f }\rangle_{\mu_M}.
\end{eqnarray}
The equality is the central result of this work. \\
\\
Proof: Since the diffusion process is Markovian, we may rewrite
the term of angle brackets on the left-hand side of
(\ref{RfunctionFeedback}) as
\begin{eqnarray}
\label{NewRfunctionFeedback}\fl\int \left(\prod_{k=M}^1dx_k
dy_k\right) R_{\lambda^M_t}(t_f|{x }_M,t_M) \prod_{k=M}^1
P(y_k|\mu_{k-1})R_{\lambda_t^{k-1}}({x
}_k,t_k|{x}_{k-1},t_{k-1})\frac{\varrho(x_k,t_k,\lambda^k_{t_k})}{\varrho(x_k,t_k,\lambda^{k-1}_{t_k})},\nonumber\\
\end{eqnarray}
where the subscript $\lambda_t^{k}$ indicates the external
parameters during the time interval $(t_k,t_{k+1})$.
Substituting~(\ref{Texplanation}) and~(\ref{GDetailbalance}) into
the above equation, we have
\begin{eqnarray}
\label{proofstep} \fl\frac{1}{\varrho({x }_0,t_0)}\int
\left(\prod_{k=1}^M d{x }_k dy_k\right) \left[\prod_{k=1}^M
P(y_k|\mu_{k-1})G_{\tilde{\lambda}_s^{k-1}}({\tilde
{x}}_{k-1},s_{k-1}|{\tilde {x }}_k,s_k)\right]
q_{\tilde{\lambda}_s^M}({\tilde{x}}_M,s_M)
\langle O\rangle_{\varrho_f}\nonumber\\
\fl=\frac{1}{\varrho({x }_0,t_0 )}\int \prod_{1}^M dy_k\prod_{1}^M
P(y_k|\mu_{k-1}) \prod_{1}^M d{\tilde{x }}_k
G_{\tilde{\lambda}_s^{k-1}}({\tilde {x }}_{k-1},s_{k-1}|{\tilde {x
}}_k,s_k) q_{\tilde{\lambda}_s^M}({\tilde{x}}_M,s_M)
\langle O\rangle_{\varrho_f}\nonumber\\
\fl=\frac{1}{\varrho({x }_0,t_0)}\int \prod_{1}^M dy_k
P(\mu_M)\langle O\rangle_{\varrho_f}
q_{\tilde{\Lambda}}({\tilde{x}}_0,t_f),
\end{eqnarray}
where $s_k$$+$$t_k$$=$$t_f$, and the subscript
$\tilde{\lambda}_s^{k}$ indicates that the transition probability
of the reversed process is under the control of the reversed
protocol.\\
\\
There are two simple observations in the above proof. First, the
$I$-term~(\ref{tjinformation}) is indispensable if one wants to
rewrite (\ref{RfunctionFeedback}) as (\ref{NewRfunctionFeedback}).
Second, extending the fluctuation relations with feedback control
essentially depends on the validity of the relations themselves.
Finally, there is a highly analogous GIFR~\cite{LiuFJPA09} in the
continuous time master equation with discrete
states~\cite{Gardiner}. Its extension of feedback control can be
easily carried out as what we did here; see the details
in~\ref{AppendixA}.

\section{Results and discussion}
\label{discussion}
\subsection{Existing special cases }\label{subsection1}
The very general~(\ref{GIFRFeedback}) in fact presents an infinite
number of equalities due to almost arbitrary choice of $\varrho$
and $S$. However, it is nontrivial to reveal some of them of
physical interest. We first show that~(\ref{GIFRFeedback}) can
cover several known fluctuation relations with feedback
control~\cite{SagawaPRL10,AbreuPRL12}. Below we set $O({x})$$=$$1$
for simplicity. The first example is for the diffusion process
that starts from the steady state $\rho_{ss}({x },\lambda^0_0)$.
If the system has instantaneous steady-state solution
$\rho_{ss}({x},\lambda^k_t)$ at any fixed parameters
$\lambda^k_t$, we may choose the solution as $\varrho$ and let
${S}$=0. Then the GIFR~(\ref{GIFRFeedback}) reduces into
\begin{eqnarray}
\label{HatanoSasaFeedback} \left\langle
e^{\sum_{k=0}^M\int_{t_k}^{t_{k+1}}
\partial_\tau \ln \rho_{ss}({x }(\tau),\tau,\lambda^k_\tau)d\tau -J-I}\right\rangle_{\rho_0} =1.
\end{eqnarray}
This is the generalized Jarzynski-Hatano-Sasa equality with
feedback control. Abreu and Seifert first presented the equality
in the discrete master equations~\cite{AbreuPRL12}. Specifically,
if the steady-state solution is just the instantaneous thermal
equilibrium state,
$\rho_{eq}({x},\lambda^k_t)$,~(\ref{HatanoSasaFeedback}) may be
further simplified into the work
equality~(\ref{orgSagawaUedaEquality}) of Sagawa and
Ueda~\cite{SagawaPRL10}. Note that under this circumstance if the
control parameters jump at the measurement times, the $J$-term
represents the amount of energy input by the external controller
into the system. In Sec.~\ref{Overdampedmodel}, we use a Brownian
model to illustrate this point.

The second example is to choose $\varrho$ to be the pdf $\rho({x
},t,\lambda^k_t)$ of the system and $S_i$ to be its irreversible
current,
\begin{eqnarray}
\label{irrersiblecurrent} J^{ir}_i(\rho)=A^{ir}_i\rho-\frac{1}{2}
\partial_{x_l} (B_{il}\rho),
\end{eqnarray}
where $A^{\rm ir}_i$ is the irreversible component of the drift
force~\cite{LiuFJPA10,Gardiner,Risken}. Then the
GIFR~(\ref{GIFRFeedback}) reduces into
\begin{eqnarray}
\label{totentropyFeedback} \left\langle
e^{-\sum_{k=0}^M\int_{t_{k}}^{t_{k+1}}{\cal J}[\rho,{J}^{\rm
ir}(\rho)](x(\tau),\tau,\lambda^k_{\tau})d\tau -J-
I}\hspace{0.1cm}\right\rangle_{\rho_0}=1.
\end{eqnarray}
We have shown that, for the nonequilibrium processes with
predetermined external parameters, the
integrand~(\ref{Jfunctional}) with these specific $\varrho$ and
${S}$ is the rate of the total entropy
production~\cite{LiuFJPA10}. Therefore, the
equality~(\ref{totentropyFeedback}) is nothing but its extension
of taking feedback control into
account~\cite{AbreuPRL12,SagawaPRL12}. The reader is reminded that
the $J$-term here is generally zero, since the pdf of the system
is continuous in time even if the control parameters have finite
jumps. However, this is no longer true for the parameters being
$\delta$-functions, {\it e.g.}, an underdamped Brownian model in
Sec.~\ref{underdampedBM}. Contrary to the
equality~(\ref{HatanoSasaFeedback}), which requires the existence
of the instantaneous steady-state
solutions,~(\ref{totentropyFeedback}) holds for general diffusion
processes.

\subsection{Generalized Bochkov-Kuzovlev equality with feedback
control} In addition to the above known results in the literature,
the GIFR~(\ref{GIFRFeedback}) predicts a previously unnoticed
equality in the dynamic perturbation problem~\cite{Risken,Kubo}.
Consider a perturbed system with the Fokker-Planck operator ${\cal
L}$$=$${\cal L}_0$$+$${\cal L}_e(t)$, where ${\cal L}_0$ denotes
the time-independent operator of the free system, and ${\cal
L}_e(t)$ is the perturbation in which the control parameters are
involved. Before applying the perturbation, the system is assumed
to be at the thermal state $\rho_{0}({x})$. Choosing $\varrho$ to
be the equilibrium pdf and setting ${S}$$=$$0$, we obtain
\begin{eqnarray}
\label{GBKFeedback} \left\langle
e^{-\sum_{k=0}^M\int_{t_{k}}^{t_{k+1}}\rho^{-1}_0{\cal
L}_e(\rho_{0}) (x(\tau),\tau,\lambda^k_{\tau})d\tau -
I}\right\rangle_{\rho_0}=1.
\end{eqnarray}
Obviously, the $J$-term here exactly vanishes. We
name~(\ref{GBKFeedback}) the generalized BKE with feedback
control. The physical relevance of above equality may be clearly
explained by the 1-d underdamped Brownian particle:
\begin{eqnarray}
\label{underdampedBPx} dq=\frac{p}{m} dt, \hspace{0.5cm}\\
\label{underdampedBPp} dp=-\partial_q {H_0}dt+\lambda^k_t
dt-\gamma\frac{p}{m}dt + \sqrt{\frac{2\gamma}{\beta}}
\hspace{0.1cm}d{W},
\end{eqnarray}
where $x$$=$$(q,p)$ is the coordinate of the particle in its phase
space, $\gamma$ is the friction coefficient, the dynamic force
$\lambda^k_t$ represents the control parameter, and
${H_0}$$=$$p^2/2m$$+$$U(q)$ is the Hamiltonian of the free system.
Note that the force may be nonconservative~\cite{Lebowitz}. Given
these notations, we immediately have $\rho_0$$\propto$$e^{-\beta
H_0(x)}$ and ${\cal L}_e$$=$$-\lambda^k_t\partial_p$. Substituting
them into~(\ref{GBKFeedback}), we have
\begin{eqnarray}
\label{BKFeedback} \left\langle
e^{-\beta\sum_{k=0}^M\int_{t_k}^{t_{k+1}} \lambda^k_{\tau} dq  -
I}\right\rangle_{\rho_0}=1.
\end{eqnarray}
We see that the integral therein is the work done by the external
force $\lambda^k_t$ on the system. In the absence of the feedback
control, (\ref{BKFeedback}) is the canonical
BKE~\cite{Bochkov77,Jarzynski07,Horowitz07}. On the other hand,
unexpectedly, for the 1-d overdamped Brownian particle, {\it
i.e.}, (\ref{underdampedBPp}) with vanishing momentum terms,
(\ref{GBKFeedback}) leads into an alternative equality
\begin{eqnarray}
\label{noncanonicalEq} \left\langle
e^{-\beta\sum^M_{k=0}\int_{t_k}^{t_{k+1}}
{\gamma}^{-1}\lambda^k_{\tau}
\partial_xU(x(\tau))d\tau -I}\right\rangle_{\rho_0}=1,
\end{eqnarray}
where $\rho_0$$\propto$$e^{-\beta U(x)}$ and ${\cal
L}_e$$=$$-\gamma^{-1}\lambda_t^k\partial_x$. We do not clearly
understand the physical meaning of the above equality, though the
integral indeed possesses the dimension of work. Even though we
must emphasize that for the overdamped particle the
BKE~(\ref{BKFeedback}) is still true but it does not originate
from (\ref{GBKFeedback}). We leave the discussion
in~\ref{AppendixB}. 

\subsection{Time reversal explanation }
Horowitz and Vaikuntanathan~\cite{HorowitzPRE10}, and Sagawa and
Ueda~\cite{SagawaPRE12} have discussed the time reversal
explanation of~(\ref{orgSagawaUedaEquality}) in great details. The
equality was thought to be the consequence of the detailed
trajectory fluctuation theorem with discrete feedback control.
This theorem is about the ratio of the joint probability of
observing trajectory and protocol in the forward process and the
joint probability of observing reversed trajectory and protocol in
the time-reversed process~\cite{JarzynskiRev11}. Different from
the forward protocols determined by the feedback measurements, the
reversed protocols are randomly drawn from the known probability
distribution of the protocols recorded in the forward
process~\cite{SagawaPRL10,HorowitzPRE10}. Interestingly, this
result automatically emerges in~(\ref{RfunctionFeedback}): its
right-hand side is the statistical average of the pdf
$q_{\tilde{\Lambda}}({\tilde{x}}_0,t_f)$ of the time reversed
process with the specific reversed protocol $\tilde{\Lambda}$ over
the pdf $P(\mu_M)$; there are no measurements at all. We must
emphasize that~(\ref{RfunctionFeedback}) is valid for very general
diffusion processes. Particularly, the content of the
time-reversal is far broader than that considered in previous
works~\cite{HorowitzPRE10,SagawaPRE12}.

In addition to the interest in the concept, the time reversal
explanation of~(\ref{RfunctionFeedback}) is also useful. We can
obtain an alternative inequality with feedback control:
\begin{eqnarray}
\label{inequalityFeedback}
\fl\left\langle\sum^M_{k=0}\int_{t_{k-1}}^{t_k}{\cal
J}[\varrho,S]({x
}(\tau),\tau,\lambda^k_\tau)d\tau\right\rangle_{\varrho_0}+\langle
J \rangle_{\varrho_0}\ge D[\varrho(x,0)||{\left\langle
q_{\tilde{\Lambda}}(\tilde{x},t_f)\right\rangle_{\mu_M}}]-\langle
I\rangle_{\varrho_0}.
\end{eqnarray}
It is worth pointing out that, the direct application of the
Jensen's inequality to~(\ref{GIFRFeedback}) leads into another
very similar inequality, where the nonnegative $D$-term above is
absent. Compared with the lower bound set by $-\langle
I\rangle_{\varrho_0}$ alone, {\it e.g.},
(\ref{SagawaIneq})~\cite{SagawaPRL10,AbreuPRL12},
(\ref{inequalityFeedback}) presents a stronger one on the
entropy-like quantity. This inequality is the other center result
of this work. In the next section, we will concretely illustrate
it by two Brownian models with single feedback control.

\subsection{Examples}\label{example}
\subsubsection{Overdamped Brownian particle}
\label{Overdampedmodel} 
\begin{figure}
\begin{center}
\includegraphics{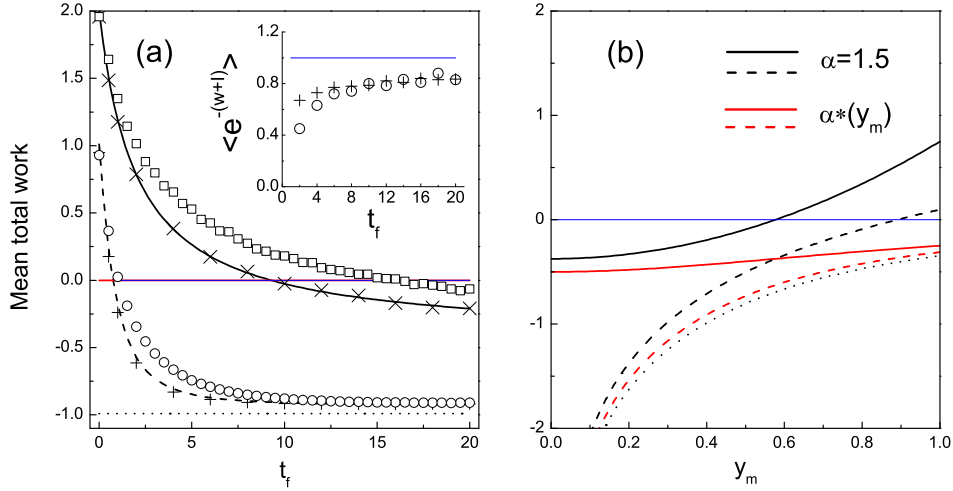}
\caption{(a). The solid, dashed, and dotted line are the mean
total work $\langle W^\star\rangle$~(\ref{optimalwork}),
$D$$-$$\langle I\rangle_{\varrho_0}$, and $-\langle
I\rangle_{\varrho_0}$, respectively, where $D$ is
(\ref{Dtermexample1}). The cross symbols are the results of the
simulation. The parameters are $y_m$$=$$0.4$ and
$\lambda_f$$=$$2$. The open symbols are the simulation data for
the case with the position-dependent friction coefficient, where
$c$$=$$0.5$. Inset: the cross and open symbols are the left-hand
side of~(\ref{HatanoSasaFeedback}) obtained by the simulation;
they are for the cases with the constant and position-dependent
friction coefficients, respectively. (b). The solid, dashed, and
dotted lines are the mean total work $\langle
W_{BK}\rangle$~(\ref{BKworkexample}), $D$$-$$\langle
I\rangle_{\varrho_0}$, and $-\langle I\rangle_{\varrho_0}$,
respectively, where $D$ is~(\ref{Dtermexample2}). The x-axis is
$y_m$. (Color online)}\label{figure1}
\end{center}
\end{figure}

The model of the overdamped Brownian
particle in~\cite{AbreuEPL11} is a good example to show the effect
of the discontinuous control parameters and to verify the
inequality~(\ref{inequalityFeedback}). At time 0 the particle is
at thermally equilibrium state $\rho_{eq}(x,0)$ in a harmonic
potential $x^2/2$. By measuring position $x$ of the particle with
outcome $x_m$, which follows a Gaussian pdf $P(x_m|x)$$=$${\cal
N}_{x_m}(x,y_m)$, the center of the potential is moved according
to a protocol $\lambda(t|x_m)$ and reaches the fixed $\lambda_f$
at the terminal time $t_f$. The equation of motion of the particle
is simply
\begin{eqnarray}
\label{overdampedBP} dx=-[x-\lambda(t|x_m)]dt + \sqrt{2}
\hspace{0.1cm}d{W}.
\end{eqnarray}
Note that all physical quantities are dimensionless. Abreu and
Seifert~\cite{AbreuEPL11} found that work can be extracted from
the single heat bath if the protocol follows the optimal one:
\begin{eqnarray}
\label{optimalprotocol}
\lambda^\star_t(x_m)=\frac{\lambda_f-b(x_m)}{t_f+2} (t+1)+b(x_m),
\end{eqnarray}
where $b(x_m)$$=$$x_m/(1+y_m^2)$. Obviously, the optimal protocol
has two jumps at time 0 and $t_f$. Since we are interested in the
work, we choose $\varrho$ to be
\begin{eqnarray}
\rho_{eq}(x,\lambda^\star_t(x_m))=\frac{1}{\sqrt{2\pi}}\exp\left[-\frac{1}{2}(x-\lambda^\star_t(x_m))^2\right],
\end{eqnarray}
and $S$$=$$0$ for~(\ref{inequalityFeedback}). Its left-hand side
is then written as
\begin{eqnarray}
\label{workcomponents} -\int dx_m P(x_m)\int dx\hspace{0.05cm}
^{x}\left\langle \int_0^{t_f}
\dot{\lambda}^\star_{\tau}(x_m)(x(\tau)-\lambda^\star_{\tau}(x_m))d\tau\right\rangle\rho(x,0|x_m)\nonumber
\\
+\int dx_m P(x_m)\int dx 
\left[\frac{1}{2}( x-\lambda^\star_0(x_m))^2-\frac{1}{2}x^2\right]
\rho(x,0|x_m)
\nonumber\\
+\int dx_m P(x_m)\int dx \rho(x,t|x_m)\left [\frac{1}{2}(
x-\lambda_f)^2-\frac{1}{2}(x-\lambda^\star_{t_f}(x_m))^2\right],
\end{eqnarray}
where $\rho(x,t|x_m)$ is the pdf of the Brownian particle at time
$t$ under the protocol $\lambda^\star_t(x_m)$, and the initial
condition is defined by
\begin{eqnarray}
P(x_m)\rho(x,0|x_m)=P(x_m|x)\rho_{eq}(x,0).
\end{eqnarray}
We clearly see that~(\ref{workcomponents}) is the total mean work
$\langle W^\star \rangle$ done on the particle~\footnote{Besides
the optimal protocol, (\ref{workcomponents}) of course holds for
arbitrary continuous protocol with two jumps at the beginning and
ending of the protocol. }: the first equation is the work done by
the continuous part of the protocol, and the latter two equations
are the energy input due to the two jumps at the beginning and
ending times. One may check that (\ref{workcomponents}) is
equivalent to Eq.~(32) in~\cite{AbreuEPL11}, {\it e.g.}, by
directly substituting~(\ref{optimalprotocol})
into~(\ref{workcomponents}) and obtaining
\begin{eqnarray}\label{optimalwork}
\langle W^\star
\rangle=\frac{\lambda_f^2}{t_f+2}-\frac{t_f}{2(t_f+2)}\frac{1}{1+y_m^2}.
\end{eqnarray}
On the other hand, given the optimal protocol, the $D$-term on the
right-hand side of~(\ref{inequalityFeedback}) is exactly
\begin{eqnarray}
\label{Dtermexample1}
D=-\frac{1}{2}\left[\ln\left(1-\frac{1}{1+K}\right)+\frac{1}{1+K}\right]
+\frac{1+K}{2K}\left(\frac{2\lambda_f}{t_f+2}\right)^2>0
\end{eqnarray}
where $K$$=$$(1+y_m^2)(1+2/t_f)^2$. The mutual information
$\langle I\rangle$ is $\ln(1+y^{-2}_m)/2$~\cite{AbreuEPL11}. In
Fig.~\ref{figure1}(a) we show~(\ref{inequalityFeedback}) at
different $t_f$ for typical $y_m$ and $\lambda_f$; see the lines
therein. We find that the presence of the $D$-term can indeed
improve the lower bound {\it on} the mean work, especially at
shorter $t_f$ in which considerable dissipation occurs. In order
to check the correctness of the analytical expressions, we also
perform the Langevian simulation under the same parameters; see
the crosses in the same figure. We see that the theory and
simulation excellently agree with each other.

In the above model we have assumed that the friction coefficient
is constant. However, in many real situations it may vary with
position~\cite{Kampen,Lancon,Dufresne,VolpePRL}, {\it e.g.},
Brownian particle near a wall. In this case, the stochastic term
in~(\ref{overdampedBP}) is multiplicative noise~\cite{Gardiner}
rather than the previously additive noise. The position-dependent
friction coefficient $\gamma(x)$ has significant influence on the
dynamics of the particle: (\ref{overdampedBP}) has to be
appropriately revised~\cite{LauPRE,YangPRE,KuroiwaJPA} as
\begin{eqnarray}
\label{overdampedBPMN} \gamma(x)dx=-[x-\lambda^\star(t|x_m)]dt
-\partial_x \gamma(x)/r(x)+\sqrt{2\gamma(x)} \hspace{0.1cm}d{W}.
\end{eqnarray}
This equation of motion generally has no analytical solutions.
According to our previous discussion,
however,~(\ref{HatanoSasaFeedback}) and
corresponding~(\ref{inequalityFeedback}) still hold
since~(\ref{overdampedBPMN}) has the instantaneous equilibrium
solution $\rho_{eq}(x,\lambda^\star_t(x_m))$. To illustrate this
claim, we assume $\gamma(x)$$=$$1+c x^2$~\cite{Lancon,YangPRE} and
simulate the mean work and the $D$-term; see the empty squares and
circles in Fig.~(\ref{figure1})(a). We find that introduction of
the position-dependent friction coefficient does not significantly
change their features under the optimal
protocol~(\ref{optimalprotocol}) and the given parameters. 
Additionally, we also note that in the current model the numerical
verification of the equality~(\ref{HatanoSasaFeedback}) is more
difficult than the verification of the inequality; see the inset
in Fig.~\ref{figure1}(a).

\subsubsection{Underdamped Brownian particle}
\label{underdampedBM} Here we use an underdamped Brownian
particle~(\ref{underdampedBPx}) and~(\ref{underdampedBPp}) with
feedback control to verify the
inequality~(\ref{inequalityFeedback}) from the BKE. The potential
is assumed to be harmonic, $U(q)$$=$$q^2/2$. At the beginning the
particle is at thermal equilibrium. After measuring velocity $v$
of the particle with outcome $v_m$ at time 0, we immediately apply
an impulse force
\begin{eqnarray}
\label{forceprotocol} \lambda_t(v_m)=-\alpha v_m\delta(t-0),
\end{eqnarray}
the strength $\alpha v_m$ of which depends on the measured
velocity. We assume that the measurement bears an Gaussian error
$P(v_m|v)$$=$${\cal N}_{v_m}(v,y_m)$. Obviously, if the
measurement is exact, {\it i.e.}, $y_m$=0, the force will make the
velocity $v$ of the particle jumping to $(1-\alpha)v$. The average
work done on the particle is
\begin{eqnarray}
\label{BKworkexample} \langle W_{BK}\rangle
=\frac{1+y_m^2}{2}\left(\alpha-\frac{1}{1+y_m^2}\right)^2-\frac{1}{2(1+y_m^2)}.
\end{eqnarray}
The work may be negative and especially has a minimal. Hence,
there exists an optimal force protocol $\lambda_t^{\star}(v_m)$
with $\alpha^\star$$=$$1/(1+y_m^2)$ that we can extract the
maximum work from the single heat bath. Additionally, the $D$-term
here has a simply expression:
\begin{eqnarray}
\label{Dtermexample2} D=\frac{1}{2}\ln[1+(1+y_m^2)\alpha^2]
-\frac{(1+y_m^2)\alpha^2}{2[1+(1+y_m^2)\alpha^2]}>0.
\end{eqnarray}
The mutual information is the same with that in the preceding
example. We show the inequality~(\ref{inequalityFeedback}) for
$\alpha$$=$$1.5$ and $\alpha^\star$ at different $y_m$ in
Fig.~1(b).

\section{Conclusion}\label{Conclusion}
In this work, we presented the GIFR with feedback control for
general diffusion processes. This general relation not only
reobtains the existing fluctuation relations with feedback control
but also predict other unnoticed relations. Moreover, we derived
an alternative inequality about the entropy-like quantity with an
improved lower bound. Two Brownian particle models are
demonstrated to verify the claim. Our discussion clearly shows
that, given any integral fluctuation relation in classical
stochastic Markovian systems, one can always obtain its
counterpart when the feedback control is taken into account. In
our opinion, hence, it is not very essential to derive these
relations one by one again in future.

\ack F.L. was supported in part by the National Science Foundation
of China under Grants No. 11174025 and Kavli Institute for
Theoretical Physics China.

\appendix
\section{GIFR with feedback control in master equations}
\label{AppendixA} Although the extension is straightforward,
considering that very different notations are involved in the
master equation, for reader's convenience we first give a review
of the GIFR~\cite{LiuFJPA09,LiuFJPA12} in this distinctive system.
We assume that the master equation has the following form:
\begin{eqnarray}
\label{masterEq} \frac{dp_n(t)}{dt}=\left[H(t)p(t)\right]_n,
\label{forwardeq}
\end{eqnarray}
where $n$ is the state index, the $N$-d column vector
$p(t)$$=$$(p_1,\cdots,p_N)^{T}$ is the probability of the system
at individual states at time $t$, and the matrix element
$(H)_{mn}$=$H_{mn}(t)$$>$0 ($m$$\neq$$n$) is the rate and
$(H)_{nn}$=$-$$\sum_{m\neq n}H_{mn}(t)$. Given an arbitrary
normalized positive column vector
${\varrho}(t)$=$(\varrho_1,\cdots,\varrho_N)^{T}$ and a
$N$$\times$$N$ matrix ${A}$ whose elements $(A)_{mn}$$=$$A_{mn}$
($m$$\neq$$n$) satisfy the conditions of
$H_{mn}\varrho_n$$+$$A_{mn}$$>0$ and $A_{nn}$$=$$-\sum_{m\neq
n}A_{mn}$, we found that there is a GIFR~\cite{LiuFJPA09}
\begin{eqnarray}
\label{MasterEqGIFT} \fl\hspace{1cm} \sum_{n=1}^N\varrho_n(t_0)
R_n(t_0)= \sum_{n=1}^N\varrho_n(t_0)^n\left\langle
e^{-\int_{t_0}^{t_f}{\cal
J}[\varrho,A](x(\tau),\tau)d\tau}O_{{x}(t)}\right\rangle= \langle
O\rangle_{\varrho_f},
\end{eqnarray}
where $O$$=$$(O_1,\cdots,O_N)$ is an arbitrary $N$-d vector,
$^{n}\langle$$\cdots$$\rangle$ is statistical average over all
trajectories starting from the state $n$ at time $t_0$, and
$\langle$~$\rangle_{\varrho_f}$ indicates an averages over
$\varrho(t_f)$. The integrand in~(\ref{MasterEqGIFT}) is
\begin{eqnarray}
{\cal
J}[\varrho,A](x(\tau),\tau)&=&\varrho_{x(\tau)}^{-1}\left[-\partial_{\tau}\varrho+
H \varrho+A 1\right]_{{x}(\tau)}+{\cal Q}[\varrho_{x(\tau)}^{-1}
A]\label{functional},
\end{eqnarray}
where the $N$-d vector $1$$=$$(1,\cdots,1)^T$,
\begin{eqnarray} {\cal
Q}[ B]&=&-B_{{x}(\tau){x}(\tau)}-\ln\left[1+\frac{B_{{x}(\tau^+){
x}(\tau^-)}(\tau)}{H_{x(\tau^+){x}(\tau^-)}(\tau)}
\right]\sum_{i=1}^K\delta(\tau-\tau_i),\label{functionalG}
\end{eqnarray}
$x(\tau)$ is the discrete state of the system at time $\tau$,
$x(\tau^{-})$ and $x(\tau^{+})$ represent the states just before
and after a jump occurring at time $\tau$, respectively, and we
assumed the jumps occur $K$ times for a trajectory. Analogous to
the GIFR~(\ref{GIFR}), the GIFR~(\ref{MasterEqGIFT}) can as well
cover several known fluctuation relations in the literature by
selecting specific ${\varrho}$ and ${A}$~\cite{LiuFJPA09}.
(\ref{MasterEqGIFT}) has a time-reversal explanation:
\begin{eqnarray}
\label{reversedprobdefinition} \varrho_n(t')R_n(t')=\langle
O\rangle_{\varrho_f}q_{\tilde{n}}(s),
\end{eqnarray}
where $q(s)$ is the $N$-d probability vector of the time-reversed
master equation,
\begin{eqnarray}
\frac{dq_{n}(s)}{ds}=[\widetilde{H}(s)
q(s)]_{n},\label{timereversal}
\end{eqnarray}
and the elements of the matrix $\widetilde{H}(s)$ are
\begin{eqnarray}
\widetilde{H}_{n m}(s)={\varrho}_{\tilde
m}^{-1}(t')\left[H_{{\tilde m}{\tilde n}}(t'){\varrho}_{\tilde
n}(t')+A_{{\tilde m}{\tilde n}}(t')\right]
\label{ratetimereversal}
\end{eqnarray}
for $m$$\neq$$n$ and $\widetilde{H}_{ mm}(s)$$=$$-\sum_{ n\neq
 m}\widetilde{H}_{nm}(s)$. Because the master equation~(\ref{masterEq}) is Markovian and
possesses equations like~(\ref{Texplanation})
and~(\ref{GDetailbalance})~\cite{LiuFJPA09}, repeating the same
derivations as those in the main text we then obtain the GIFR with
feedback control in the master equation:
\begin{eqnarray}
\label{GIFT} \left\langle
e^{-\sum_{k=0}^M\int_{t_k}^{t_{k+1}}{\cal J}[\varrho,A]({x
}(\tau),\tau,\lambda^k_\tau)
-J-I}O_{{x}(t)}\right\rangle_{\varrho_0}=\langle \langle
O\rangle_{\varrho_f} \rangle_{\mu_M},
\end{eqnarray}
where
\begin{eqnarray} \label{jumpprotocolsME}
J=\ln\prod_{k=1}^{M}\frac{
\varrho_{x(t_k)}(t_k,\lambda^{k-1}_{t_k})}{\varrho_{x(t_k)}(t_k,\lambda^{k}_{t_k})},
\end{eqnarray}
and the $I$-term is still~(\ref{tjinformation}). Here we have
assumed that the rates $H_{mn}(t)$ depend on time only through the
external control parameters $\lambda_t$ and feedback control is
completely the same with that in the diffusion processes; see
Sec.~\ref{GIFRfeedbacksection}.

\section{BKE of overdamped Brownian particle}
\label{AppendixB} Here we only discuss the fluctuation relation
without feedback control. Specifically choosing
$\varrho$$=$$\rho_0$ and $S$$=$$\rho_0 \lambda(t)/\gamma$ and
substituting them into the GIFR~(\ref{GIFR}), we can obtain the
canonical BKE~\cite{Bochkov77} for the overdamped Brownian
particle, where we used the notation $\lambda(t)$ instead of
$\lambda^k_t$. On the basis of~(\ref{reversedforce})
and~(\ref{reverseddiffusion}), we find that the BKE
and~(\ref{noncanonicalEq}) have distinctive time reversal
interpretations. For the former the time-reversed SDE is the same
with the forward one except that the dynamic force is changed into
$\lambda(t_f-t)$, whereas for the latter the direction of the
force is also reversed, namely, $-\lambda(t_f-t)$. This point can
be explicitly checked for the Brownian particle in the harmonic
potential $U(x)$$=$$x^2/2$. After a simple calculation, we find
that before integrating on $x_0$ the left-hand side of
(\ref{noncanonicalEq}) is
\begin{eqnarray}
\label{testR} \fl\hspace{1cm}^{x_0}\langle
\hspace{0.1cm}e^{-\int_0^{t_f}
\lambda(\tau)x(\tau)d\tau}\rangle=\exp\left[-x_0\int_0^{t_f}
\lambda(\tau)e^{-\tau}d\tau-\frac{1}{2}\left(\int_0^{t_f}\lambda(\tau)e^{-\tau}\right)^2\right].
\end{eqnarray}
Multiplying~(\ref{testR}) by the thermal state
$\rho(x_0,0)$$\propto$$e^{-U(x_0)}$, we obtain
\begin{eqnarray}
^{x_0}\langle \hspace{0.1cm}e^{-\int_0^{t_f}
\lambda(\tau)x(\tau)d\tau}\rangle \rho(x_0,0)={\cal
N}_{x_0}\left(-\int_0^{t_f} \lambda(\tau)e^{-\tau}d\tau,1\right),
\end{eqnarray}
which is just the pdf $q(x_0,t_f)$ of the time-reversed process
with the dynamic force $-\lambda(t_f-t)$. On the other hand,
interestingly, one may check that the time reversal explanation of
the BKE for the underdamped Brownian
particle,~(\ref{underdampedBPx}) and~(\ref{underdampedBPp}),
agrees with that of the BKE for the overdamped Brownian particle.
These little surprising characteristics of the overdamped Brownian
particle are due to lack of the degree of freedom of velocity.


\section*{References}
\begin{thebibliography}{99}

\bibitem{Evans}
Evans D J Cohen E G D and Morriss G P 1993 {\it Phys. Rev. Lett.}
{\bf 71} 2401


\bibitem{EvansSearles} Evans D J and Searles D J 1994 {\it
    Phys. Rev. E} {\bf 50} 1645


\bibitem{Gallavotti} Gallavotti G and Cohen E G D 1995 {\it
    Phys. Rev. Lett.} {\bf 74} 2694

\bibitem{Kurchan}
Kurchan J 1998 {\it J. Phys.} A: {\it Math. Gen.} {\bf 31} 3719

\bibitem{Lebowitz}
Lebowitz J L and Spohn H 1999 {\it J. Stat. Phys.} {\bf 95} 333

\bibitem{Bochkov77}
Bochkov G N and Kuzovlev Yu E 1977 {\it Sov. Phys. JETP} {\bf 45}
125

\bibitem{JarzynskiPRL97}
Jarzynski C 1997 {\it Phys. Rev. Lett.} {\bf 78} 2690

\bibitem{JarzynskiPRE97}
Jarzynski C 1997 {\it Phys. Rev. }E {\bf 56} 5018


\bibitem{Crooks99}
Crooks G E  1999 {\it Phys. Rev. }E {\bf 60} 2721


\bibitem{Crooks00}
Crooks G E  1999 {\it Phys. Rev. }E {\bf 61} 2361

\bibitem{HatanoSasa}
Hatano T and Sasa S I 2001 {\it Phys. Rev. Lett.} {\bf 86} 3463

\bibitem{Maes}
Maes C 2003 {\it Sem. Poincare} {\bf 2} 29

\bibitem{SeifertPRL05}
Seifert U 2005 {\it Phys. Rev. Lett.} {\bf 95} 040602


\bibitem{Speck} Speck T and Seifert U 2005 {\it J. Phys.} A:
    {\it Math. Gen.} {\bf 38} L581

\bibitem{Kawai} Kawai R Parrondo J M R and Van den Broeck C
    2007 {\it Phys. Rev. Lett.} {\bf 98} 080602

\bibitem{EspositoPRL10} Esposito M and Van den Broeck C 2010
    {\it Phys. Rev. Lett.} {\bf 104} 090601

\bibitem{Leff} Leff H S and Rex A F, eds. 2003 {\it Maxwell��s
Demon 2: Entropy, Classical and Quantum Information, Computing}
(Princeton: Princeton University Press)

\bibitem{CaoEntropy12} Cao F J and Feito M 2012 {\it Entropy} {\bf
    14} 834

\bibitem{SagawaPRL10} Sagawa T and Ueda M 2010 {\it Phys. Rev.
    Lett.} {\bf 104} 090602

\bibitem{CaoPRE09} Cao F J and Feito M 2009 {\it Phys. Rev. E}
    {\bf 79} 041118

\bibitem{Toyabe} Toyabe S Sagawa T Ueda M Muneyuki E and Sano M 2010 {\it Nat. Phys.} {\bf 6} 988

 \bibitem{AbreuPRL12} Abreu D and Seifert U 2012 {\it Phys.
    Rev. Lett.} {\bf 108} 030601

\bibitem{KunduPRE12} Kundu A 2012 {\it Phys. Rev. E} {\bf 86}
021107

\bibitem{LahiriJPA12} Lahiri S Rana S and Jayannavar A M 2012 {\it
    J. Phys. A: Math. Theor.} {\bf 45} 065002

\bibitem{SagawaPRL12} Sagawa T and Ueda M 2012 {\it Phys. Rev.
Lett} {\bf 109} 180602

\bibitem{Chernyak} Chernyak V Chertkov M and Jarzynski C 2006
    {\it J. Stat. Mech: Theor. Exp.} P08001

\bibitem{Taniguchi} Taniguchi T and Cohen E G D 2008 {\it J
Stat Phys} {\bf 130} 633

\bibitem{Chetrite} Chetrite R and Gawedzki K 2008 {\it
    Commun. Math. Phys.} {\bf 282} 469

\bibitem{LiuFPRE09} Liu F and Ou-Yang Z C 2009  {\it Phys. Rev.} E
    {\bf 79} 060107(R)

\bibitem{LiuFJPA09} Liu F Luo Y P Huang M C and Ou-Yang Z C
    2009 {\it J. Phys. A: Math. Gen.} {\bf 37} 332003

\bibitem{LiuFJPA10} Liu F Tong H Ma R and
    Ou-Yang Z C 2010 {\it J. Phys. A: Math. Theor.} 43 495003

\bibitem{HorowitzPRE10} Horowitz J and Vaikuntanathan S 2010
    {\it Phys. Rev. E} {\bf 82} 061120

\bibitem{SagawaPRE12} Sagawa T and Ueda M 2012
    {\it Phys. Rev. E} {\bf 85} 021104

\bibitem{Gardiner} Gardiner C W 1983 {\it Handbook of
    stochastic methods} (Berlin: Springer).

\bibitem{Risken}
Risken H 1984 {\it The Fokker-Planck equation} (Berlin: Springer)

\bibitem{Cover} Cover T M and Thomas J A 2006 {\it Elements of Information
Theory} (New York: Wiley-Interscience)

\bibitem{JarzynskiRev11} Jarzynski C 2011 {\it Annu. Rev. Condens. Matter Phys.} {\bf 2}
329

\bibitem{FeitoPRE09} Feito M Baltanas J P and Cao F J 2009 {\it Phys. Rev. E} {\bf 80}
031128

\bibitem{LopezPRL08} Lopez B J Kuwada N J Craig E M Long B R and
Linke H 2008 {\it Phys. Rev. Lett} {\bf 101} 220601

\bibitem{AbreuEPL11} Abreu D and Seifert U 2011 {\it EPL} {\bf 94} 10001

\bibitem{Kubo}
Kubo R 1966 {\it Rep. Prog. Phys.} {\bf 29} 255

\bibitem{Jarzynski07}
Jarzynski C 2007 {\it C. R. Phys.} {\bf 8} 495

\bibitem{Horowitz07} Horowitz J and  Jazynski C 2007 {\it J. Stat. Mech:
Theory Exp.} P11002

\bibitem{Kampen} van Kampen N 1988 {\it J. phys. Chem. Solids}
{\bf 49} 673

\bibitem{Lancon} Lancon P Batrouni G Lobry L and Ostrowsky N 2001 {\it
EPL} {\bf 54} 28

\bibitem{Dufresne} Dufresne E R Altman D and Grier D G 2001 {\it EPL} {\bf 53}
264

\bibitem{VolpePRL} Volpe G Helden L Brettschneider T Wehr J and Bechinger C
2013 {\it Phys. Rev. Lett.} {\bf 104} 170602

\bibitem{LauPRE} Lau A W C and Lubensky T C
2007 {\it Phys. Rev.} E {\bf 76} 011123

\bibitem{YangPRE} Yang M and Ripoll M
2013 {\it Phys. Rev.} E {\bf 87} 062110

\bibitem{KuroiwaJPA} Kuroiwa T and Miyazaki K
2014 {\it J. Phys. A: Math. Gen.} {\bf 76} 011123

\bibitem{LiuFJPA12} Liu F and Hong L 2012 {\it J. Phys. A: Math.
Gen.} {\bf 45} 125004

\end{thebibliography}
\end{document}